\documentclass[RNAAS,twocolumn]{aastex63}

\usepackage{color, xspace}

\received{December 4, 2020}
\accepted{December 5, 2020}
\published{December 9, 2020}

\shorttitle{Vanilla long term integrations of the Solar System}
\shortauthors{Brown et al.}

\begin{document}

\title{A repository of vanilla long term integrations of the Solar System}

\correspondingauthor{Hanno Rein}
\email{hanno.rein@utoronto.ca}

\author[0000-0002-9354-3551]{Garett Brown}
\affiliation{Department of Physical and Environmental Sciences, University of Toronto at Scarborough, Toronto, Ontario M1C 1A4, Canada}
\affiliation{Department of Physics, University of Toronto, Toronto, Ontario, M5S 3H4, Canada}

\author[0000-0003-1927-731X]{Hanno Rein}
\affiliation{Department of Physical and Environmental Sciences, University of Toronto at Scarborough, Toronto, Ontario M1C 1A4, Canada}
\affiliation{Department of Physics, University of Toronto, Toronto, Ontario, M5S 3H4, Canada}
\affiliation{Department of Astronomy and Astrophysics, University of Toronto, Toronto, Ontario, M5S 3H4, Canada}


\begin{abstract}
    We share the source code and a 121~GB dataset of 96 long-term N-body simulations of the Solar System. 
    This dataset can be analyzed by itself to study the dynamics of the Solar System.
    In addition, our simulations can be the starting point for future studies wishing to explore different initial conditions or additional physical effects on the Solar System.
    Our simulations can also be used as a comparison and benchmark for new numerical algorithms.
\end{abstract}

\section{Motivation} \label{sec:intro}
The long term dynamical evolution of the Solar System has been investigated analytically for centuries.
However, direct numerical integrations of the Solar System over billions of years have only become possible in the last few decades thanks to fast computers and better numerical algorithms.
Since then, many authors have run such simulations, including \cite{Applegate1986}, \cite{BatyginLaughlin2008}, and \cite{LaskarGastineau2009}.
These simulations differ in the initial conditions and numerical methods used, as well as in the physical effects that are being modelled.
For example, some simulations only include gravitational forces, while others also include stellar mass loss or tidal forces.

The Solar System is chaotic and even small changes in the initial conditions lead to rapidly diverging trajectories. 
For this reason, one typically has to integrate many simulations with slightly different initial conditions and make statistical conclusions.
Each such simulation is computationally expensive and takes of the order of a month of wall-time to complete a 5~Gyr integration even on modern computers. 
Because $N$-body simulations are inherently sequential for small $N$, it is practically impossible to parallelize them. 
The best one can do is run one simulation on one core, but this still requires a minimum wall-time of one month, even if a large number of cores is available.

We share a set of 96 long-term integrations of the Solar System for which the average relative energy error is $\Delta E/E\sim 2.5\cdot10^{-9}$.
However, our aim is not to generate the most accurate ephermeris possible, but to provide a repository of vanilla simulations that capture the most important dynamical properties of the system.

We hope that these vanilla simulations will allow other studies to 1) analyze this dataset by itself, and 2) run their own simulations and use this dataset as a comparison. 
To facilitate this, we not only share the data, but also all the code used to generate it. 
Our simulations can therefore be used as a benchmark and reference in future studies exploring additional physical effects, additional bodies, different initial conditions, or different numerical algorithms.

\section{Numerical setup} \label{sec:setup}
We share all scripts to exactly reproduce our simulations together with our dataset.
In this section, we only summarize the important aspects of the numerical scheme used.

We use the $N$-body code \texttt{REBOUND} \citep{ReinLiu2012} and the symplectic integrator \texttt{WHFast} \citep{WisdomHolman1991, ReinTamayo2015} in
Jacobi coordinates \citep{ReinTamayo2019}.
To achieve a high accuracy we employ 17th order symplectic correctors \citep{Wisdom2006} and a modified kick step \citep{Wisdom1996,2019MNRAS.489.4632R}.
Specifically, we use the \textit{lazy implementer's kernel method} which is compatible with general relativistic corrections.
This configuration corresponds to using the shorthand \texttt{WHCKL} for the integrator in \texttt{REBOUND}, having a generalized order (18,4,3), and leading error terms $O(\epsilon dt^{18}+\epsilon^2 dt^4 + \epsilon^3 dt^3)$.
For more details on various high order symplectic schemes in \texttt{REBOUND} see \cite{2019MNRAS.489.4632R}.
We use a fixed timestep of $dt\approx8.062$~days.
This implies that our simulations can be trusted as long as the eccentricity of any planet remains moderate,~$e\lesssim 0.4$. 

We query NASA's Horizon system\footnote{\url{https://ssd.jpl.nasa.gov/?horizons}} to retrieve initial conditions of the Sun and all 8 major planets on 1st of January 2000, 12:00 UTC.
The effects of general relativistic precession are important for the long-term evolution of the Solar System \citep{Laskar2008}.
We model these effects by including an additional non-Newtonian $1/r^3$ term in the potential using the \texttt{gr\_potential} module of \texttt{REBOUNDx} \citep{Tamayo2019}. 
Our simulations use a system of units where the unit of length is one astronomical unit, the unit of mass is one Solar mass, and the gravitational constant~$G$ is~$1$.
One Earth year then corresponds to~$2\pi$ in code units.

All 96 simulations are identical except for a small perturbation in the $x$ coordinate of Mercury.
The file names indicate by how much Mercury has been shifted. 
For example, in the simulation labelled \texttt{p750}, Mercury has been shifted by $750\cdot0.38$mm in the positive $x$ direction. 
Note that as long as the remain small the details of these initial perturbations are not important because the system is chaotic.

\begin{figure}[t]
\includegraphics[width=0.99\columnwidth]{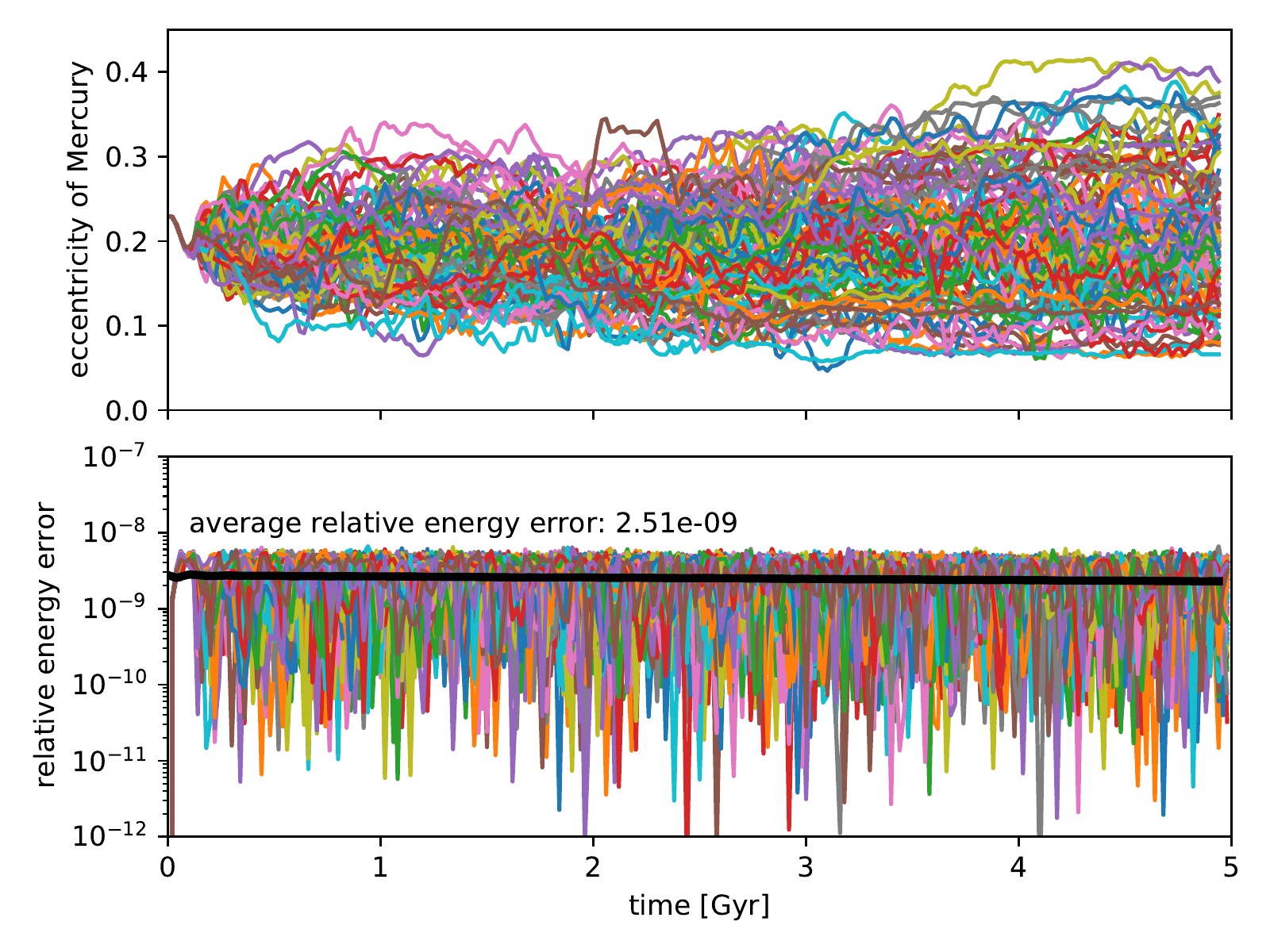}
\caption{Mercury's eccentricity (top) and the relative energy error (bottom) as a function of time. Showing all 96 simulations.  \label{fig:eccentricity}}
\end{figure}

The simulations are integrated for 5~Gyr into the future. 
We make use of the \texttt{SimulationArchive} \citep{ReinTamayo2017} to store 500,000 snapshots for each simulation, resulting in a file size of 1.3~GB.
These snapshots can be used to sample a simulation at 10,000 year intervals.
Furthermore, the simulations can be restarted and resampled at any snapshot. 
The trajectories will be bit-by-bit identical to the original integration.

\section{Data access} \label{sec:data}
The 96 SimulationArchive files as well as all our scripts are made available on the general-purpose open-access repository Zenodo at \url{https://zenodo.org/record/4299102} \citep{brown_2020_4299102}.
The full dataset is over 121~GB in size but each SimulationArchive can also be downloaded as an individual 1.3~GB file.

There are two different ways to think about a SimulationArchive.
It can simply be used to access the data stored at each of the 500,000 snapshots. 
However, one can also think of each snapshot as a point where one can restart the simulation. 
Because two consecutive snapshots are only 3s apart in wall-time, one can access any arbitrary time in the 5~Gyr integration within at most 3s (1.5s on average).
And because the SimulationArchive allows the simulation to be reconstructed bit-by-bit, this is not an interpolation but exact repetition of the original simulation.

Fig.~\ref{fig:eccentricity} shows the eccentricity of Mercury for all 96 integrations in the top panel and the relative energy error in the bottom panel. 
Even though we need to sift through 121~GB of data, generating this plot only takes about one minute on a desktop computer with a solid state drive. 
We can see that the eccentricity of Mercury is correlated across all simulations for the first $\sim100$~Myrs.
All simulations remain stable for the full 5~Gyr which is consistent with previous results showing that only about 1\% of simulations should go unstable \citep{LaskarGastineau2009}.
The bottom panel shows that the relative energy error remains approximately constant throughout the integration. 
The average relative energy error of our simulations is about $2.5\cdot 10^{-9}$ and is shown as a thick black line in the bottom panel.
The Zenodo repository includes example scripts to query the SimulationArchive files and reproduce the above figure as well as a few others.

The dataset we share took about 6 core years of computation time on a 2.4GHz Intel Xeon CPU, resulting in 200kg of $\text{CO}_2$ emissions\footnote{Assuming 0.5kg $\text{CO}_2$ per~kWh, \url{https://www.eia.gov/tools/faqs/faq.php?id=74}.}.
This research was made possible by the open-source projects 
\texttt{Jupyter} \citep{jupyter}, \texttt{iPython} \citep{ipython}, 
and \texttt{matplotlib} \citep{matplotlib, matplotlib2}.

\bibliography{main}{}
\bibliographystyle{aasjournal}

\end{document}